\documentclass[final,3p,times,twocolumn]{elsarticle}

 \usepackage{graphics}

\usepackage{amssymb}

\journal{Solid State Communications}

\begin{document}

\begin{frontmatter}



\title{Quantum Hall activation gaps in bilayer graphene}

 \author[a1]{E.~V.~Kurganova\corref{cor1}}
 \cortext[cor1]{Corresponding author: tel.: +31243653052, fax: +31243652440}
 \ead{E.Kurganova@science.ru.nl}
 \author[a1]{A.~J.~M.~Giesbers}
 \author[a2]{R.~V.~Gorbachev}
 \author[a2]{A.~K.~Geim}
 \author[a2]{K.~S.~Novoselov}
 \author[a1]{J.~C.~Maan}
 \author[a1]{U.~Zeitler}
 
 \address[a1]{High Field Magnet Laboratory, Institute for Molecules and Materials,
Radboud University Nijmegen, Toernooiveld 7, 6525 ED Nijmegen, The Netherlands}
 \address[a2]{School of Physics and Astronomy, University of Manchester, M13 9PL, Manchester, United Kingdom}



\begin{abstract}
 We have measured the quantum Hall activation gaps in bilayer graphene at filling factors $\nu=\pm4$ and $\nu=\pm8$ in high magnetic fields up to 30~T. We find that energy levels can be described by a 4-band relativistic hyperbolic dispersion. The Landau level width is found to contain a field independent background due to an intrinsic level broadening and a component which increases linearly with magnetic field.

\end{abstract}

\begin{keyword}

A. Bilayer graphene \sep D. Quantum Hall effect \sep D. Landau levels 

\PACS \sep 71.70.Di \sep 73.43.-f \sep 72.80.Vp

\end{keyword}

\end{frontmatter}

\newpage


\section{Introduction}
\label{}
Bilayer graphene consists of two coupled one atom thick carbon layers arranged in a honeycomb lattice. These two layers are coupled in an A1-B2 stacking, where A and B represent the two triangular sublattices of the individual graphene layers ~\cite{CastroNeto,Falko}, see inset in Fig.~\ref{QH}.

Bilayer graphene attracted the attention of researchers in transistor electronics due to its tunable gap induced by an electric field ~\cite{Castro,VandersypenNature,DubnaNano} or by doping ~\cite{Ohta}.
At the same time its hyperbolic dispersion of massive chiral Dirac fermions \cite{Geim2009} makes it an intriguing object for more fundamental physics as a connecting link between the 2D single layer graphene with relativistic massless charge carriers and the 3D classical Fermi gas in graphite.

In particular, placing bilayer graphene in a magnetic field leads to an unconventional integer quantum Hall effect \cite{Falko,BiUli} distinctively different from the relativistic half-integer quantum Hall effect in single layer graphene \cite{KostyaPioner,KimPioner} and the integer quantum Hall effect in the conventional 2D electron systems \cite{IQH}.
 
In this Communication we probe the hyperbolic dispersion of bilayer graphene by measuring the activation gaps between neighboring Landau levels.
We will show that the activation energies are described by the Landau level spectrum for a 4-band relativistic hyperbolic dispersion with a Landau level width composed of two components, namely a field independent intrinsic broadening and a linearly field dependent broadening.

\section{Experimental results and discussion}
 \label{}

\begin{figure}[t]
  \vspace*{2em}
  \begin{center}
  \includegraphics[width=0.95\linewidth]{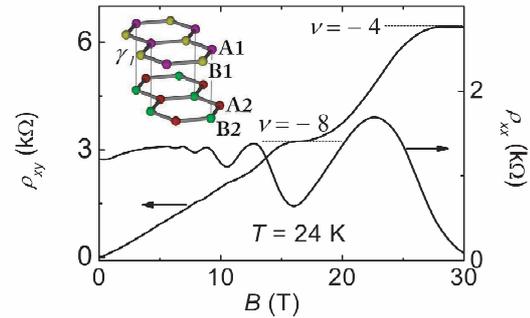}
  \end{center}
\vspace*{0em}
\caption{(color online). Hall resistivity $\rho_{xy}$ and longitudinal  resistivity $\rho_{xx}$ as a function of magnetic field at $T=24$ K for a hole-doped device ($n=-3.07\times10^{16}$ m$^{-2}$ ). The inset shows the A1-B2 stacking in bilayer graphene.}
\label{QH}
\end{figure}

We have measured the resistivity of a 1 $\mu$m wide Hall bar of bilayer graphene as a function of magnetic field up to 30 T. The sample was fabricated by micromechanical exfoliation of a flake from natural graphite, deposited on a 300 nm SIMOX wafer and identified using an optical microscope ~\cite{Kostya2004}. The carrier concentration in such samples can be changed by applying voltage to the underlying conducting silicon, which acts as a back gate. 
To remove surface impurities we have annealed the sample at 390 K in vacuum prior to measurements, giving on an increasing the mobility from $\mu~=~1000$~cm$^{2}$/(Vs) to $\mu~=~3000$~cm$^{2}$/(Vs).

Figure ~\ref{QH} shows typical traces of the Hall resistivity $\rho_{xy}$ and the longitudinal resistivity $\rho_{xx}$ for a hole-doped sample (charge carrier concentration $n=-3.1\times10^{16}$ m$^{-2}$) as a function of magnetic field at a temperature of 24 K. 
The Hall resistivity exhibits plateaus at filling factors $\nu=-4i$ where $i=1,2,3,\dots$ and the ``$-$'' sign indicates holes. These plateaus are accompanied by minima in the longitudinal resistivity.

\begin{figure}[t]
  \begin{center}
  \includegraphics[width=0.95\linewidth,angle=0]{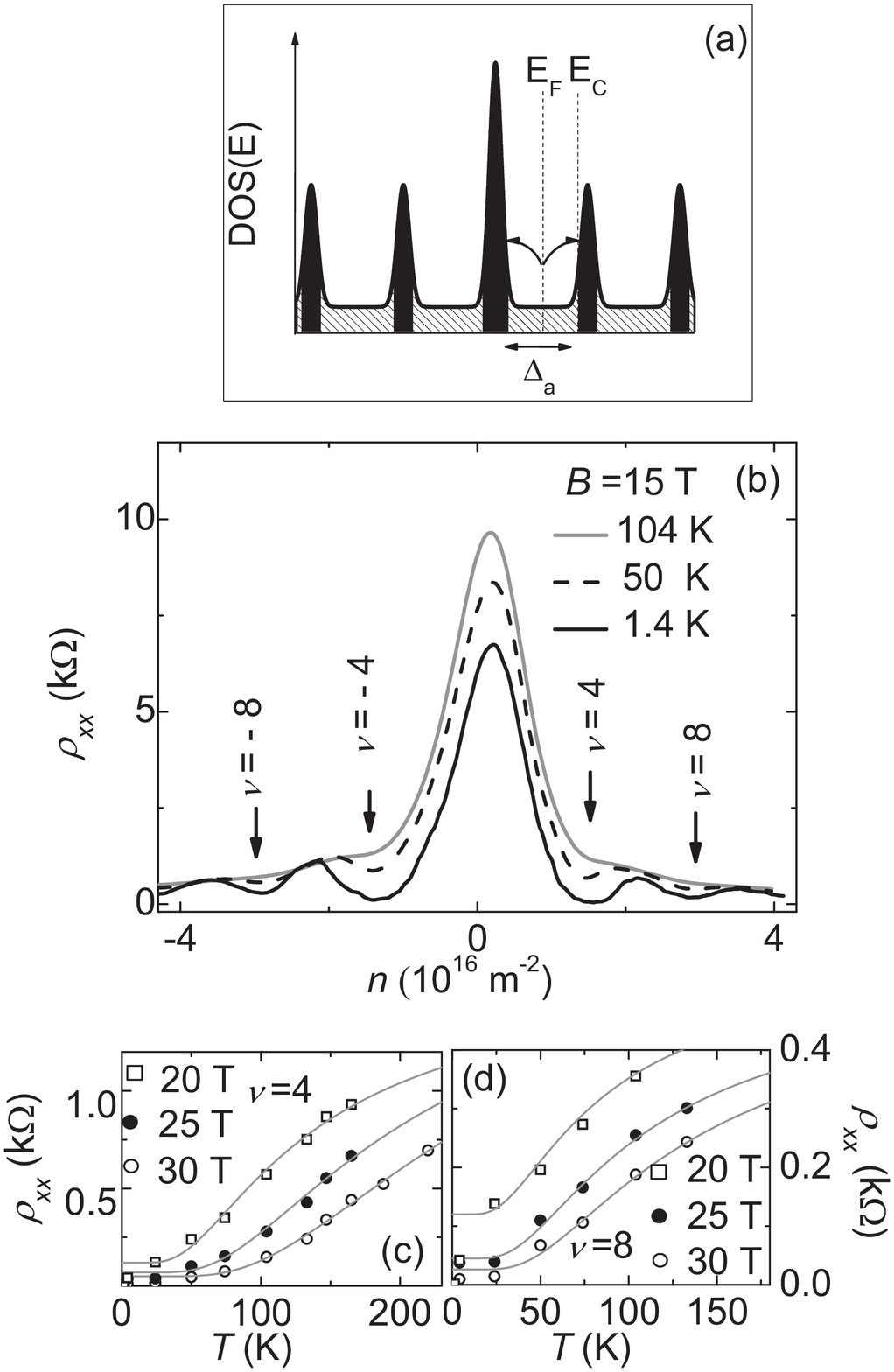}
  \end{center}
  \vspace*{0em}
  \caption{Density of states modelled with extended (black areas) and localised (dashed areas) states (a). Longitudinal resistivity $\rho_{xx}$ as a function of charge carrier concentration at $B=15$ T and various temperatures (b). Temperature dependences of $\rho_{xx}$ at various fields at $\nu =4$ (c) and $\nu =8$ (d). The solid lines represent a fit, using the Fermi-Dirac distribution function.}
  \label{Fermifits}
\end{figure}

The appearance of quantum Hall plateaus at $\nu=\pm4i$ can be understood from the Landau level spectrum of bilayer graphene shown in Fig.~\ref{Fermifits}(a). The Landau level at the charge neutrality point is 8-fold degenerate consisting of 4 electron and 4 hole levels. The other Landau levels are each 4-fold degenerate and have a hole (``$-$'' sign) or electron (``$+$'' sign) character. The density of states of each Landau level consists of a region of extended states, which carry current, near the center, while states in the upper or lower tails are localised. Minima of $\rho_{xx}$ occur when the Fermi energy (which can be varied by sweeping gate voltage or field) lies in the middle between two Landau levels. The minima, corresponding to $\nu=\pm4$ and $\nu=\pm8$, are shown in Fig.~\ref{Fermifits}(b) by straight arrows.

To determine activation gaps we measured the longitudinal resistivity $\rho_{xx}$ as a function of charge carriers concentration $n$ at various temperatures $T=0.5$~K$-220$~K Fig.~\ref{Fermifits}(b). At the highest field used, Shubnikov de Haas oscillations remain visible up to 220~K. The resistivity at the minima remains zero up to 24 K. This temperature is considerably low compared with single layer graphene, where the quantum Hall effect can even be observed up to room temperature ~\cite{RTQH}. 

With increasing temperature electrons can be thermally activated from localised states to the extended states of the neighboring Landau levels (arrows in the Fig.~\ref{Fermifits}(a)) increasing $\rho_{xx}$ at a minimum.
The temperature dependences of $\rho_{xx}$ for $\nu=4$ and $\nu=8$ are shown in Fig.~\ref{Fermifits}(c) and (d) (at $\nu=-4$ and $\nu=-8$ similar results  are obtained). For low temperatures the dependence of $\rho_{xx}$ on temperature flattens off significantly, which is commonly attributed to non-activated transport mechanism as variable range hopping \cite{VonKlitzing}.
The gray lines in Fig.~\ref{Fermifits}(c), (d) show a fit of the experimental $\rho_{xx}(T)$ dependences by Fermi-Dirac function $1/(1+exp(\Delta_{a}/2k_{B}T))$ to describe the activated transport. The activation gaps $\Delta_{a}$ are used as a fit parameter.

We will show below that the activation gaps measured do not exceed the thermal energy considerably. Therefore we use the Fermi-Dirac function to extract the gaps, rather than the commonly used low temperature exponential approximation (Arrhenius plot).
We assume that the conductivity is determined by the thermal activation to the single energy between localised and extended states, namely the conductivity edge $E_{C}$ of the Landau levels (Fig.~\ref{Fermifits}(a)). The thermal activation deeper inside the Landau levels is neglected. We have verified the validity of this assumption by modelling the longitudinal resistivity $\rho_{xx}$ inside a gap $\Delta_{a}$ between two broadened Landau levels using standard Kubo-Greenwood formalism ~\cite{Kubo,Greenwood}.
Indeed, for $\Delta_{a}>2k_{b}T$ the gaps given by this model agree within 2\% with those extracted from the Fermi-Dirac fit.

\begin{figure}[t]
  \begin{center}
  \includegraphics[width=0.8\linewidth,angle=0]{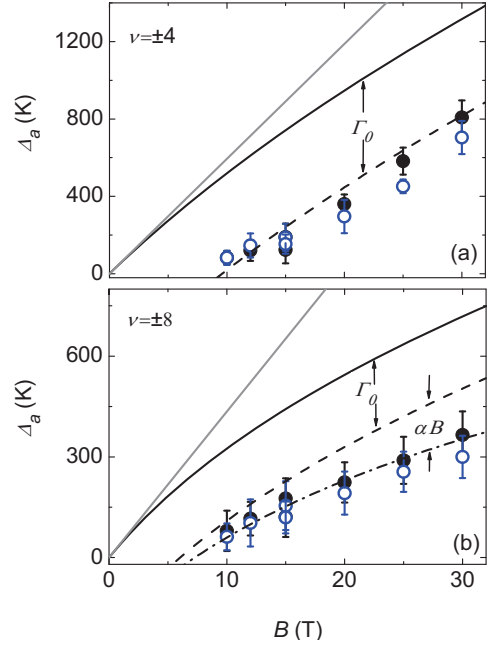}
  \end{center}
  \vspace*{0em}
\caption{(color online). The activation gaps $\Delta_{a}$ extracted from the fit (Fig.~\ref{Fermifits}(c), (d)) as a function of magnetic field at  $\nu =\pm4$ (a) and $\nu =\pm8$ (b). Black filled (blue empty) circles correspond to electrons (holes). The black solid lines show the distances between centers of corresponding Landau levels as given by hyperbolic dispersion Eq.~(\ref{LL}), the gray solid lines represent the parabolic approximation Eq.(\ref{FalkoLL}). The dashed lines represent the expected field dependence for a constant level broadening, the dash-dotted line includes a field dependent level width (see the main text for details).}
\label{gaps}
\end{figure}

The experimental gaps, extracted from the Fermi-Dirac fit are presented in Fig.~\ref{gaps}.
Due to the flattening of experimental $\rho_{xx}(T)$ curves  at low temperatures ($T<70$ K for $\nu=\pm4$ and $T<50$ K for $\nu=\pm8$) we limited our fitting range to higher temperatures but included a non-activated small constant background. Extending the fitting to lower temperature yields systematically different gaps. We indicated this effect by the size of the error bars in Fig.~\ref{gaps}.
 
The activation gap determined this way corresponds to the double energy from the Fermi energy to the nearest extended state and is therefore given by $\Delta_{a}=\Delta E-\Gamma$, with a Landau level broadening $\Gamma$ and an energy $\Delta E$ between the centers of the levels . The measured energy is thus lower than the bare gap of infinitely sharp Landau levels and even vanishes when $\Gamma=\Delta E$.

Theoretically the low band Landau levels in bilayer graphene are given by ~\cite{Pereira}:
\begin{eqnarray}
\label{LL}
E_{N}&=&\frac{{\rm sgn}(N)} {\sqrt{2}}
         \Bigl[(2|N|+1)2eBc^{2}\hbar+\gamma_{1}^{2}\\
     &-&   \sqrt{\gamma_{1}^{4}+4(2|N|+1)eBc^{2}\hbar\gamma_{1}^{2} 
                 +(2eBc^{2}\hbar)^{2}}\Bigr]^{1/2} 
\nonumber
\end{eqnarray}

with $N=0,\pm1,\pm2\dots$, the ``$\pm$'' sign corresponds to electrons or holes.
In this equation $\gamma_{1}=0.4$ eV \cite{JunYan,Li} is the interlayer coupling constant and $c=10^{6}$ ms$^{-1}$ ~\cite{KostyaPioner} is the velocity in single layer graphene.
For low magnetic field Eq.~(\ref{LL}) reduces to
~\cite{Falko}:
\begin{equation}\label{FalkoLL}
 E_{N}= {\rm sgn}(N)\frac{\hbar eB}{m^{*}}\sqrt{|N|(|N|+1)}
\end{equation}
where the effective mass arises from the interlayer coupling $\gamma_{1}$ as $m^{*}=\gamma_{1}/(2c^{2})$. The low field expression Eq.~(\ref{FalkoLL}) predicts a linear field dependence, arising from a parabolic dispersion, with different slopes depending on $N$. The full Eq. (\ref{LL}) also displays a non-parabolicity for high fields and/or energies. 

In Fig.~\ref{gaps} we compare our experimental data with the calculated energy splitting for $\nu=\pm4$ and $\nu=\pm8$ using Eqs. (\ref{LL}) and (\ref{FalkoLL}). From the figure it can be seen that for the same magnetic field the experimental gaps for the $\nu=\pm8$ ($N=1\rightarrow2$) are always smaller than that for $\nu=\pm4$ ($N=0\rightarrow1$) as theoretically predicted. Furthermore the activation gaps seem to have a sublinear field dependence which is more pronounced for $\nu=\pm8$ and which is partially attributed to the predicted non-parabolicity (Eq.~(\ref{LL})). The experimental gaps vanish at a finite field below 10 T which we attribute to the vanishing of the activation gaps when broadened Landau levels start to overlap, i. e. when  $\Gamma=\Delta E$. For that reason, the measured activation gaps follow the bare gaps with increasing field with an offset that is determined by $\Gamma$. This $\Gamma$ appears to be a constant $\Gamma_{0}$ for the $\nu=\pm4$ transition (dashed line in Fig.~\ref{gaps}(a)) whereas it shows a field dependence $\Gamma=\Gamma_{0}+\alpha B$ for $\nu=\pm8$ transition (dashed-dotted and dashed lines in Fig.~\ref{gaps}(b)). This field dependent broadening for the higher Landau levels ($N\neq0$) could be caused by the rippled nature of the graphene, which leads to a difference of the perpendicular component of the magnetic field throughout the sample and therefore to an inhomogeneous linearly dependent broadening. This field dependent mechanism was also observed for non-zero energy Landau level in single layer graphene ~\cite{Giesbersgaps}. The lowest Landau level ($N=0$) is not field dependent and thus the $N=0\rightarrow1$ transition is much less affected by this mechanism.

It is interesting to see, that for bilayer graphene our measurements show a broadened zeroth Landau level up to 30 T, without any indication of a narrowing as was seen in single layer graphene ~\cite{Giesbersgaps}. It is possible that we need samples of a better quality or higher magnetic fields to observe this narrowing in bilayer graphene. It might also be an indication of a different broadening mechanism of the lowest Landau level in the single and the bilayer graphene and therefore different conditions for the coexistence of holes and electrons at the charge neutrality point.

 \section{Conclusions}
 \label{}
 
To conclude, we have investigated the gaps between Landau levels at filling factors $\nu=\pm4$ and $\nu=\pm8$ in bilayer graphene by means of activated transport measurements. Experimental field dependences follow Landau level quantization for a 4-band hyperbolic relativistic dispersion with field dependent offsets caused by a field dependent level width.

Part of this work has been supported by EuroMagNET under EU contract 228043 and by the Stichting Fundamenteel Onderzoek der Materie (FOM) with financial support from the Nederlandse Organisatie voor Wetenschappelijk Onderzoek (NWO). The work was also supported by Engineering and Physical Sciences Research Council (UK), the Royal Society, the European Research Council (programs "Ideas", call: ERC-2007-StG and "New and Emerging Science and Technology", project "Structural Information of Biological Molecules at Atomic Resolution"), Office of Naval Research, and Air Force Office of Scientific Research. The authors are grateful to Nacional de Grafite for supplying high quality crystals of graphite.

\clearpage

\end{document}